\documentclass[a4paper,fleqn]{cas-dc}

\usepackage{tikz}
\usetikzlibrary{arrows.meta, positioning, calc}
\usepackage[authoryear,longnamesfirst,round]{natbib}
\usepackage{graphicx} 
\usepackage{tcolorbox} 
\usepackage{listings} 
\usepackage{caption} 
\usepackage[utf8]{inputenc}

\begin{document}
\sloppy
\let\WriteBookmarks\relax
\def\floatpagepagefraction{1}
\def\textpagefraction{.001}
\shorttitle{Generative AI for Gambling Behavior Health}
\shortauthors{Konrad Samsel et~al.}

\title [mode = title]{Multimodal Generative AI and Foundation Models for Behavioural Health in Online Gambling}

\author[1]{Konrad Samsel}
\cormark[1]

\credit{Conceptualization of this study, Methodology, Software}

\affiliation[1]{organization={Institute of Health Policy, Management and Evaluation, Dalla Lana School of Public Health, University of Toronto},
                city={Toronto},
                state={Ontario},
                country={Canada}}

\author[1]{Mohammad Noaeen}[style=]
\cormark[1]

\author[1,2]{Neil Seeman}[%
   role=,
   suffix=,
   ]

\credit{Data curation, Writing - Original draft preparation}

\affiliation[2]{organization={Massey College, University of Toronto},
                city={Toronto},
                state= {Ontario},
                country={Canada}}

\author[1]{Karim Keshavjee}


\author[3]{Li-Jia Li}
\affiliation[3]{
organization= {LiveX AI},
City= {San Francisco},
state= {California},
country= {United States}
}
\author[1,4,5]{Zahra Shakeri}
\affiliation[4]{organization={Schwartz Reisman Institute, University of Toronto}, 
                city={Toronto},
                state={Ontario}, 
                country={Canada}}
\affiliation[5]{organization={Faculty of Information, University of Toronto},
                city={Toronto},
                state={Ontario}, 
                country={Canada}}
\cormark[2]

\cortext[cor1]{These authors contributed equally to this work.}
\cortext[cor2]{Corresponding author}

\begin{abstract}
Online gambling platforms have transformed the gambling landscape, offering unprecedented accessibility and personalized experiences. However, these same characteristics have increased the risk of gambling-related harm, affecting individuals, families, and communities. Structural factors, including targeted marketing, shifting social norms, and gaps in regulation, further complicate the challenge. This narrative review examines how artificial intelligence, particularly multimodal generative models and foundation technologies, can address these issues by supporting prevention, early identification, and harm-reduction efforts. We detail applications such as synthetic data generation to overcome research barriers, customized interventions to guide safer behaviors, gamified tools to support recovery, and scenario modeling to inform effective policies. Throughout, we emphasize the importance of safeguarding privacy and ensuring that technological advances are responsibly aligned with public health objectives.
\end{abstract}



\maketitle

\section{Introduction}

The advent of in-person electronic gambling machines (EGMs) introduced immersive experiences for players, allowing wagers to be placed as rapidly as once every 2.5 seconds \citep{dowling2005electronic}. These principles now extend to online gambling, where fast-paced digital platforms heighten the risk of behavioral addiction and related harms \citep{gainsbury2015online}. This escalation is facilitated by the unprecedented accessibility and personalized interfaces of many gambling sites, factors that can further exacerbate public health challenges tied to disordered gambling \citep{gainsbury2015online}.

Online gambling has become a major global concern, propelled by rapid technological growth and evolving social norms. In 2023, the worldwide online gambling market was valued at \$92.9 billion and is projected to grow at a compound annual rate of 11.7\% to reach \$145.6 billion by 2028 \citep{marketreport2023}. Younger populations appear especially drawn to digital platforms, with reports indicating that nearly 50\% of gamblers aged 18--34 prefer online options over traditional venues \citep{gamblingtrends2022}. This surge has an escalating public health toll, as an estimated 1--3\% of adults worldwide experience disordered gambling, affecting tens of millions of individuals \citep{who2019gambling}. The wider societal costs are likewise striking. In the United States, gambling-related problems are estimated to cost \$6 billion per year in healthcare, criminal justice, and lost productivity \citep{costofgambling2022}, while the United Kingdom, where 0.7\% of adults are classified as problem gamblers, sees a related societal expense exceeding \pounds1.2 billion annually \citep{gamblingcommission2021}.

Growth in online gambling stems not only from technological advances but also from changes in social norms and regulatory environments, creating novel challenges for public health practitioners. While online offerings parallel in-person betting in terms of game types and prizes, they possess distinctive features argued to amplify their appeal \citep{hubert2018comparison}. Surveys point to constant accessibility, anonymity, and mobile integration as influential drivers of engagement \citep{hubert2018comparison, columb2018national}, while user-friendly financial transactions and immersive interfaces may further intensify addictive tendencies \citep{gainsbury2015online}. Gambling behaviors can span a continuum from recreational to at-risk to problematic \citep{latvala2019public}, suggesting a need for proactive interventions, such as user feedback, limit-setting, self-exclusion measures, and mental health support, to mitigate escalation \citep{rodda2022systematic}.

Data analytics and artificial intelligence (AI) present innovative avenues for tackling these issues. Past research has examined AI-driven methods for detecting risky behaviors \citep{auer2023using} and supporting in-game limit-setting \citep{auer2022predicting}. Lessons from AI applications in patient care, such as psychiatric referral management \citep{habicht2024closing} and personalized treatment \citep{d2020ai}, indicate how similar frameworks might be adapted to address online gambling addictions. Nonetheless, AI can also aggravate negative outcomes if misapplied. Operators increasingly leverage advanced data analytics and risk management solutions to make platforms appear more rewarding, which can inadvertently deepen gambling-related harm \citep{tyler2023ai}. The unchecked personalization behind these immersive environments risks creating cycles of overexposure and escalating problematic behaviors.

Multimodal Generative AI and foundation models offer a new frontier for addressing these multifaceted challenges. Through real-time analytics, personalized interventions, and comprehensive policy simulations, these emerging technologies hold potential for improving early detection and harm reduction in online gambling. However, their responsible deployment requires robust governance to avoid exacerbating existing harms or introducing new risks to vulnerable populations.
In this narrative review, we synthesize the contexts under which gambling-related addictive behaviors arise and discuss how advanced AI technologies may strengthen primary prevention, early screening, and addiction management in online gambling. We also highlight ethical considerations and oversight mechanisms needed to ensure these innovations align with public health goals. Finally, we propose concrete strategies for achieving a paradigm shift in AI-driven harm reduction and detail a set of potentially high-impact applications enabled by multimodal generative AI, while underscoring the core challenges, ethical, technical, and organizational, that must be addressed to realize their full promise. 


\section{Understanding Problem Gambling}

Problem gambling, also referred to as pathological gambling or gambling disorder, is a behavioural addiction characterized by \textquotesingle{}\textit{frequent, repeated episodes of gambling that dominate the patient's life to the detriment of social, occupational, material, and family values and commitments}\textquotesingle{} \citep{who2016icd10}. Though standardized diagnostic criteria are available, the condition is often underdiagnosed due to the low proportion of individuals experiencing harmful effects who are willing to seek support \citep{lupo2023gambling}. It has been estimated that around 71\% of those with problematic gambling have never sought treatment for their condition \citep{suurvali2008treatment}. Problematic gambling behaviour, if left untreated, can be associated with harms including bankruptcy, housing instability, and worsening mental health \citep{latvala2019public}. In some cases, problem gambling can also impair judgment and decision-making abilities, leading to more impulsive and risky behaviours \citep{vestergaard2023comorbidity}. Importantly, the negative impacts associated with this condition can extend beyond individuals, often also affecting those around them \citep{tulloch2021effect}. These wider harms can include increased household financial strain, relationship breakdowns, emotional distress among affected family members, and an overall decrease in one's quality of life \citep{latvala2019public}.

Recently, Internet gaming disorder (IGD) and problematic Internet use have been identified as conditions associated with problematic gambling \citep{karlsson2019associations}. While it has been proposed that these behavioural addictions may share similar risk factors, further work is needed to determine whether any causal associations exist between them \citep{karlsson2019associations}.  The distinction between gaming and gambling is becoming increasingly blurred, with some games incorporating gambling-like elements, and gambling sites increasingly \textquotesingle{}gamifying\textquotesingle{} core features of their platforms \citep{wu2018cognitive, raneri2022role}.  The recent classification of IGD in the Fifth Edition of The Diagnostic and Statistical Manual of Mental Illnesses (DSM-V) has helped in establishing several distinctions between disordered gambling and gaming \citep{darvesh2020exploring}. Differentiating between these behavioural addictions can also assist in the development of tailored approaches to mitigating their severity and impact. While problematic gamblers tend to experience more financial-related harms, IGD is more often characterized by physical and health-related harms associated with recurrent gaming behaviour \citep{delfabbro2021harm}.

Approaches to aid in the screening of problem gambling have previously been proposed, and serve as tools to identify individuals requiring further intervention or diagnostic assessment. These criteria typically consist of readily quantifiable factors such as the frequency and duration of gambling activities and the amount of money spent on gambling \citep{gooding2023harm, turner2018cross, williams2023etiology, auer2023using}. One example of a screening tool is the Problem and Pathological Gambling Measure (PPGM-R), a 15-item self-reported assessment for classifying the risk of gambling disorder \citep{gooding2023harm}. In tandem with these screening tools, knowledge of high-risk demographics, gender-based differences, and other behavioural factors may be effective in informing public health approaches to screening and early management.

\begin{figure}
\centering
    \includegraphics[width=\linewidth]{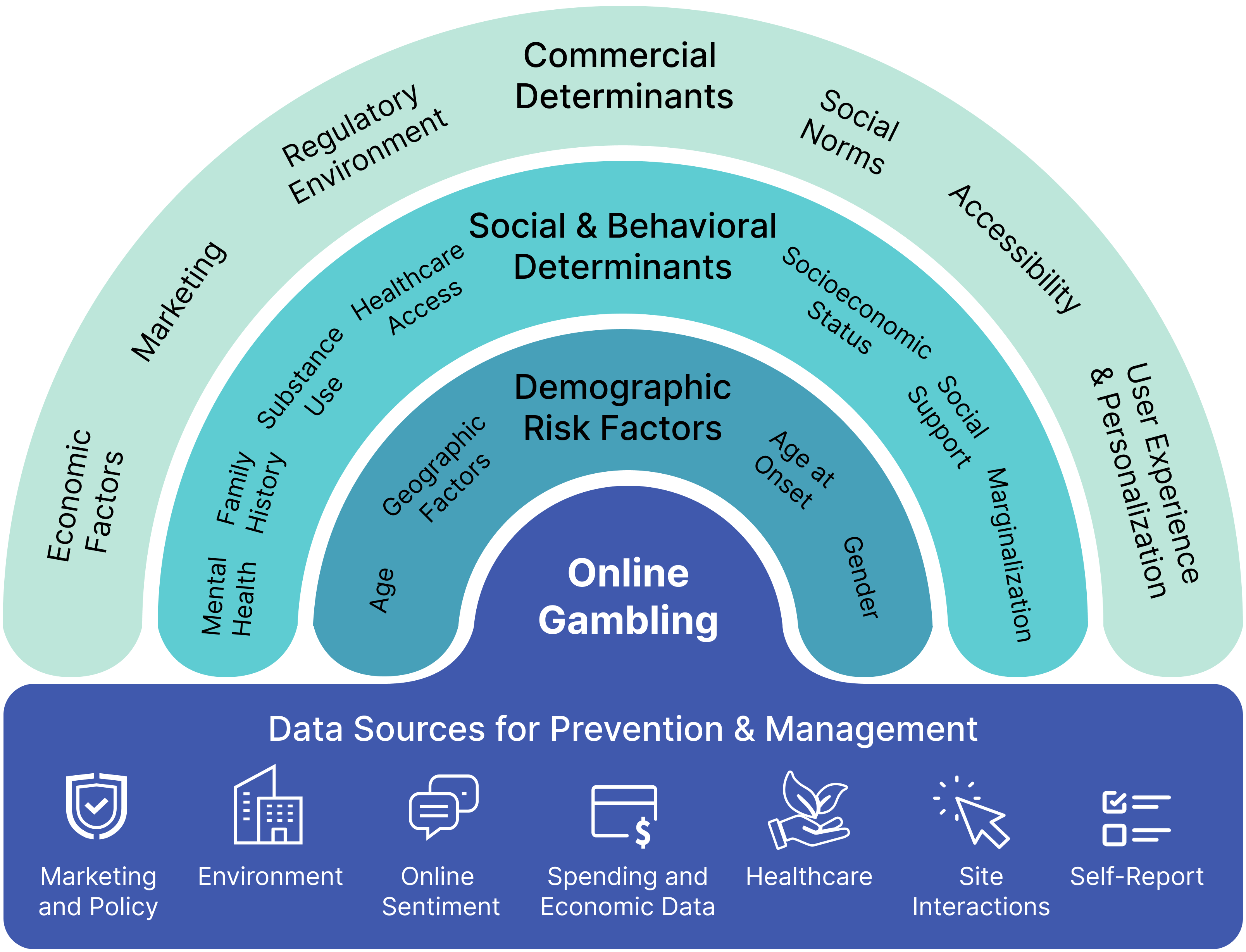}
    \caption{\textbf{Environmental determinants of online gambling and associated data sources for research and prevention.} The determinants are classified into accessibility, socioeconomic factors, regulatory environments, and commercial influences. These determinants interact to shape online gambling behaviors and outcomes. The figure also maps relevant data sources, including demographic surveys, public health records, and behavioral datasets, to support research on these influences and inform interventions aimed at reducing gambling-related harms.}
    \label{fig:determinants_and_data}
\end{figure}

\subsection{Demographic Risk Factors}

The risk of engaging in gambling behaviour has previously been associated with demographic factors including age, initial age of gambling, and gender. These factors are situated within a larger scope of social, behavioural, and environmental determinants ({Figure \ref{fig:determinants_and_data}}). Understanding risk factors associated with problem gambling is essential for designing effective interventions and prevention strategies tailored to specific populations. 

Adolescents and younger adults represent a particularly vulnerable population, being prone to engaging in risky gambling behaviours during their formative years \citep{bastiani2023childhood, rizzo2023wanna}. Adolescents' developing cognitive control systems may hinder effective decision-making in these contexts, making them more susceptible to developing problematic gambling behaviours \citep{emond2020gambling}. The age at which adolescents begin to engage in these activities has been suggested as a risk factor, with one cross-sectional study finding severe gamblers were more likely to have started gambling at an earlier age \citep{sharman2019psychosocial}. The growing inclusion of gambling-related elements in adolescent games, such as virtual wagers and loot boxes, underscores the need for research into the impacts of these early exposures to gambling tendencies in later life. Among other demographic factors, the prevalence of problem gambling has historically been higher in males \citep{williams2021predictors, turner2018cross, claesdotter-knutsson2021changes}. Online gambling, specifically, carries a similar demographic risk profile, with males and younger adults more likely to engage with these platforms \citep{hing2017risk}. 

\subsection{Social, Behavioural, and Health Determinants}

The co-occurrence of alcohol and substance use with gambling behaviours significantly increases the risk of developing gambling addiction \citep{hilbrecht2020conceptual}. Prior research has suggested that individuals who engage in excessive alcohol consumption or substance use are more likely to exhibit impulsive behaviours, including problem gambling \citep{yau2015gambling}.  The link between gambling and substance use may stem from similar neurobiological and psychological factors underlying these conditions \citep{pettorruso2021transition}. The combination of alcohol or substance use with gambling may lead to a cycle of addictive behaviours, exacerbating the development and severity of gambling addictions \citep{yau2015gambling}. The underlying mechanisms of addiction, including neurobiological \citep{potenza2011neuroscience}, emotional \citep{jauregui2016pathological}, and structural factors \citep{barnes2015gambling}, can predispose individuals to engage in multiple addictive behaviours, including gambling. Additionally, individuals with addiction to other substances may turn to gambling as an alternative or complementary form of seeking pleasure or relief \citep{potenza2011neuroscience}, increasing the likelihood of experiencing gambling-related problems \citep{armoon2023global, barnes2015gambling}. 

Problematic gambling has also been noted to co-occur with mental health diagnoses such as depression, anxiety disorders, and personality disorders \citep{hilbrecht2020conceptual}. The co-occurrence of mental health disorders and problem gambling presents complex challenges for diagnosis and treatment, emphasizing the importance of integrated approaches that address both gambling behaviours and underlying mental health conditions. In tandem, lower socioeconomic status (SES) is known to be associated with the overall incidence of problem gambling \citep{williams2021predictors, williams2023etiology}, mental health and substance use \citep{lasserre2022socioeconomic}. Interestingly, an opposite association has been observed in some studies of online gambling, where those with higher incomes and educational levels may be more likely to partake in betting activities and exhibit problematic behavioural tendencies \citep{gainsbury2015online}. It has also been argued that these differences between online and offline offerings may become less pronounced as online gambling platforms become more well-known and easier to access \citep{gainsbury2015online}. By extension, social mobility and changes in SES, predictive of mental health \citep{barakat2023review}, present an interesting domain that may benefit from investigation in an online gambling context. 

Other determinants suggested for their relationship with problematic gambling include financial literacy, family history of gambling, and access to health services. A previous investigation found financial literacy to be associated with gambling, with lower levels of financial literacy corresponding to an elevated risk of unhealthy gambling behaviours \citep{wang2023characteristics}. Moreover, family history of gambling has been proposed as an additional predictive factor of addictive gambling behaviour \citep{williams2023etiology, gooding2023harm}. Former investigations have found that individuals self-reporting perceived problematic gambling behaviours were more likely to have a family history of gambling compared to those not reporting such behaviours \citep{gooding2023individual}. Structural barriers to accessing relevant health services and a lack of social support networks can also serve as risk factors for the progression of this behavioural addiction \citep{dkabrowska2017barriers}.

\subsection{Commercial Determinants}

Commercial determinants of health present an interesting aspect of study in the context of gambling promotion and severity. Reflecting aspects such as the broad commercial landscape \citep{reith2022framing}, regulatory environments \citep{engebo2021regulatory, turner2023brief}, accessibility \citep{russell2023electronic}, product design \citep{mcauliffe2021responsible}, and broader economic conditions \citep{olason2017economic}, these determinants are increasingly being recognized as important drivers of gambling behaviour. Marketing strategies, which serve to increase awareness and positively shape perceptions on gambling, can be argued to influence social norms and help normalize engagement in these platforms \citep{guillou2021gambling}. Recent attention has also focused on the role of celebrities and social media influencers in the appeal of these platforms \citep{pitt2024young}. With this marketing often occurring on digital social media platforms, the risk of adolescents and youth being exposed to gambling messaging is also heightened \citep{guillou2021gambling}.

\section{Online Gambling Characteristics}

Developments in gambling offerings have introduced novel methods of participation, such as in-person EGMs and internet-based games. As a 'continuous' form of gambling \citep{williams2021predictors}, EGMs are characterized by their repetitive nature, which translates to a shorter recovery period in between gambling engagement and reward \citep{harris2021relationship}. Moreover, the increased prominence of auditory and visual stimuli commonly found in these machines further promotes continued user engagement \citep{shao2013shifts}. Recently, a natural policy experiment from Australia highlighted the effects of EGM access restrictions on the prevalence of gambling-related harms \citep{russell2023electronic}. Lower levels of harm were associated with access restrictions \citep{russell2023electronic}, highlighting the contributions of accessibility in gambling prevention and management. Unfortunately, the ease of access to gambling activities has tended to increase due to the emergence and worldwide growth of online gambling. 

The gambling landscape has undergone significant transformations in recent years, with the proliferation of online gambling platforms contributing to shifts in gambling behaviour and overall prevalence rates of problematic gambling behaviours \citep{marionneau2023gambling, emond2020gambling, elton2016examination, miles2023nationwide, langham2016understanding}. There is growing evidence that online gamblers are at a greater risk of experiencing problem gambling compared to those who engage in offline gambling \citep{scholes2012relationships, griffiths2011internet}.  Overall, the popularity, lack of restrictions, and personalization among online offerings are among the factors proposed to increase the likelihood of problem gambling.

{\bf Increasing Popularity of Online Gambling.} Online gambling offerings, including online wager platforms, sports gambling apps, and virtual slot machines, have witnessed an increase in popularity, particularly among adolescents \citep{emond2020gambling, gainsbury2015online}. According to a 2016 cross-sectional survey of 10,035 Canadian adolescents, 41.6\% of respondents reported engaging in gambling activities within the prior three months \citep{elton2016examination}. Moreover, based on survey data collected using the Canadian Adolescent Gambling Inventory (CAGI), it was estimated that 17.4\% and 18.2\% of those who had self-reported engaging in online gambling activities were classified at a \textquotesingle{}high\textquotesingle{} and \textquotesingle{}low to moderate\textquotesingle{} level of gambling severity, respectively \citep{elton2016examination}. Comparable estimates for youth who reported engaging in offline gambling yielded estimates of 1.2\% for \textquotesingle{}high\textquotesingle{} and 7.2\% for \textquotesingle{}low to moderate\textquotesingle{} severity \citep{elton2016examination}. Online sports betting emerged as the most popular form of online gambling among adolescents \citep{elton2016examination}. From this same survey, simulated gambling such as free online poker and gambling games on social media were also prevalent among Canadian youth, with approximately 9\% of respondents reporting past engagement in these platforms \citep{elton2016examination}. 

{\bf Ease of Access.} Online gambling platforms have shaped the gambling experience by providing unparalleled accessibility to users \citep{ghelfi2023online, emond2020gambling, marionneau2023gambling}. In contrast to traditional brick-and-mortar casinos, which include age restrictions and have limitations based on operating hours and physical proximity, online platforms allow individuals to participate in gambling activities from the comfort of their own homes or while on-the-go via mobile devices \citep{hilbrecht2020conceptual, favieri2023portrait, allami2023predictors}. Consequently, individuals may find themselves more inclined to participate in these activities due to the  accessibility afforded by online platforms \citep{marionneau2023gambling}.
  
{\bf Lack of Sufficient Age Restrictions.} One concerning aspect of online gambling platforms is the lack of stringent age verification measures compared to traditional brick-and-mortar casinos. While physical casinos typically enforce strict age restrictions and require identification verification to prevent underage gambling, online platforms may have less robust age verification mechanisms in place \citep{bastiani2023childhood}. This lax enforcement increases the risk of underage individuals engaging in gambling activities, as they may circumvent age restrictions and access online gambling platforms without adequate age verification checks \citep{elton2016examination}.

{\bf Personalization of Gambling Experience.} Online gambling platforms leverage sophisticated and personalized data analytics to tailor gambling experiences to individual preferences \citep{tyler2023ai}. By analyzing user data and behavioural patterns, these platforms can customize the presentation of games, offers, and incentives to align with each user's interests and preferences \citep{marionneau2023gambling, miles2023nationwide}. This personalized approach may enhance user engagement and immersion in the gambling experience. Moreover, integrating personalized features, such as targeted advertisements \citep{cbc2022onlinecasino} and recommendations \citep{Sahota2024forbes} based on past behaviour, may further incentivize individuals to continue gambling and increase their participation in gambling activities, particularly when presented with tailored incentives and stimuli that resonate with their preferences.

{\bf Increased Intensity.} Online gambling platforms facilitate a higher intensity, frequency, and prevalence of gambling activities compared to traditional venue-based gambling \citep{marionneau2023gambling}. Features such as unlimited pay-to-play options, minimal wait times between bets, and the availability of multiple gambling options may contribute to the heightened intensity and frequency of online gambling experiences. The absence of physical and frequency constraints, and the seamless transition between different gambling activities, enable individuals to engage in continuous and uninterrupted sessions, leading to increased intensity and frequency of problematic gambling behaviours \citep{allami2021meta-analysis, ghelfi2023online}. Compared to other gambling forms, the amount of time between bet placement and outcomes is much shorter, which may influence an individual's ability to manage impulsive tendencies \citep{ghelfi2023online, hilbrecht2020conceptual, calado2016problem}.

Online gambling has evolved into a dynamic and highly engaging ecosystem, designed to captivate and retain users. For vulnerable groups such as adolescents, the combination of instant accessibility and hyper-personalization is not merely appealing, it is dangerous. Addressing these challenges requires more than regulation; it necessitates a comprehensive understanding of the psychological drivers and structural factors underpinning this digital phenomenon, laying the groundwork for innovative and effective public health interventions. 

\section{Artificial Intelligence Addressing Problem Gambling}

Artificial Intelligence holds significant promise for addressing the complex issue of problem gambling, but it also raises key ethical questions regarding fairness and transparency. This section draws on a diverse body of research across several key areas—\textit{Detection and Prediction}, \textit{Behavioral Analysis and Risk Factors}, \textit{AI for Intervention and Prevention}, \textit{Ethical and Fairness Considerations}, {\it Preventive Strategies}, and \textit{Limitations in AI for Problem Gambling}—to provide a broad synthesis of current knowledge and challenges in this evolving field.

\subsection{Core Applications of AI in Problem Gambling}

\medskip

\noindent\textbf{Detection and Prediction.}
A substantial portion of AI research in problem gambling focuses on developing predictive models that flag at-risk individuals early. Studies have explored diverse machine learning approaches, including logistic regression, random forests, and artificial neural networks, to analyze betting transactions, session durations, and user correspondence with customer service \citep{murch2024establishing,smith2024automatic,jach2024identification,murch2023using,murch2024comparing,kairouz2023enabling,percy2016need,percy2016predicting,finkenwirth2021using,percy2020lessons,auer2023using}. Some of these models have shown lasting stability over months, with only minor threshold adjustments required to maintain accuracy \citep{murch2024comparing,murch2024establishing}. In parallel, AI chatbots have been trialed to deliver cognitive behavioral therapies, offering a scalable way to connect individuals experiencing gambling problems to early support \citep{murch2024establishing}. This trend toward automated detection and screening suggests a broader move from reactive to proactive healthcare strategies.

\medskip

\noindent\textbf{Behavioral Analysis and Risk Factors.}
Complementing predictive algorithms, a growing body of research is dedicated to understanding \emph{which behaviors} best indicate risk. Studies highlight that problem gamblers typically deposit more frequently, lose more money per session, and exhibit wider fluctuations in betting stakes \citep{emond2020gambling,auer2023using,ghaharian2024ai}. Further, analyses of online forum posts and chat data suggest a correlation between high community engagement and heightened problem gambling behaviors \citep{smith2024automatic}. However, these findings often remain correlational rather than strictly causal, reflecting the multifaceted nature of gambling harm \citep{deng2019applying,ghaharian2023applications}. Detecting behavioral patterns, such as sudden increases in deposit amounts, enables researchers and operators to improve machine learning models and create more accurate methods for identifying individuals at risk.

\medskip

\noindent\textbf{AI for Intervention}
AI is also being leveraged for \emph{intervention and monitoring} efforts, extending beyond detection to deliver personalized harm-reduction materials. Systems can recommend contacting local helplines or provide immediate feedback that contextualizes a user’s actual gambling behaviors in real-time \citep{murch2024establishing,kairouz2023enabling,sandor2024unmasking,kim2024ai,peres2021time}. Some platforms employ nudges to encourage self-imposed deposit limits, while others focus on more specialized use cases, such as detecting match-fixing in sports \citep{kim2024ai}. These interventions align with a broader push toward public health models that proactively support vulnerable individuals. Incorporating short \textquotesingle{}motivational check-ins\textquotesingle{}, tailored to a user’s recent betting trajectory, may strengthen engagement and steer players toward safer gambling habits.

\medskip

\noindent\textbf{Ethical and Fairness Considerations.}
Despite these advances, the ethical dimensions of AI in problem gambling remain complex. Concerns include the potential for algorithms to perpetuate existing inequalities if, for instance, they perform poorly for specific demographic groups \citep{murch2024comparing,percy2020lessons,ghaharian2024ai}. Debates around fairness (equality vs.\ equity) further complicate how limited resources, such as free counseling sessions, are offered to different user segments \citep{murch2024comparing}. Moreover, critiques exist that AI interventions may serve as \textquotesingle{}power mechanisms\textquotesingle{} shaping individuals into \textquotesingle{}responsible gamblers\textquotesingle{} according to industry-driven norms, rather than addressing deeper structural factors \citep{ghaharian2024ai}. Transparent communication about algorithmic limitations and an inclusive approach to model validation are crucial to mitigate such risks.

\medskip


Despite these challenges and limitations in current AI applications for problem gambling, advancements in \textbf{generative AI and foundation models} offer a promising new direction. By enabling more adaptive, scalable, and context-aware solutions, these technologies can address gaps in data quality, fairness, and personalized interventions. In the following section, we illustrate how Generative AI and foundation models can push beyond traditional approaches, enhancing research, prevention, and treatment strategies while maintaining robust ethical safeguards.

\begin{figure*}
\centering
    \includegraphics[width=.85\textwidth]{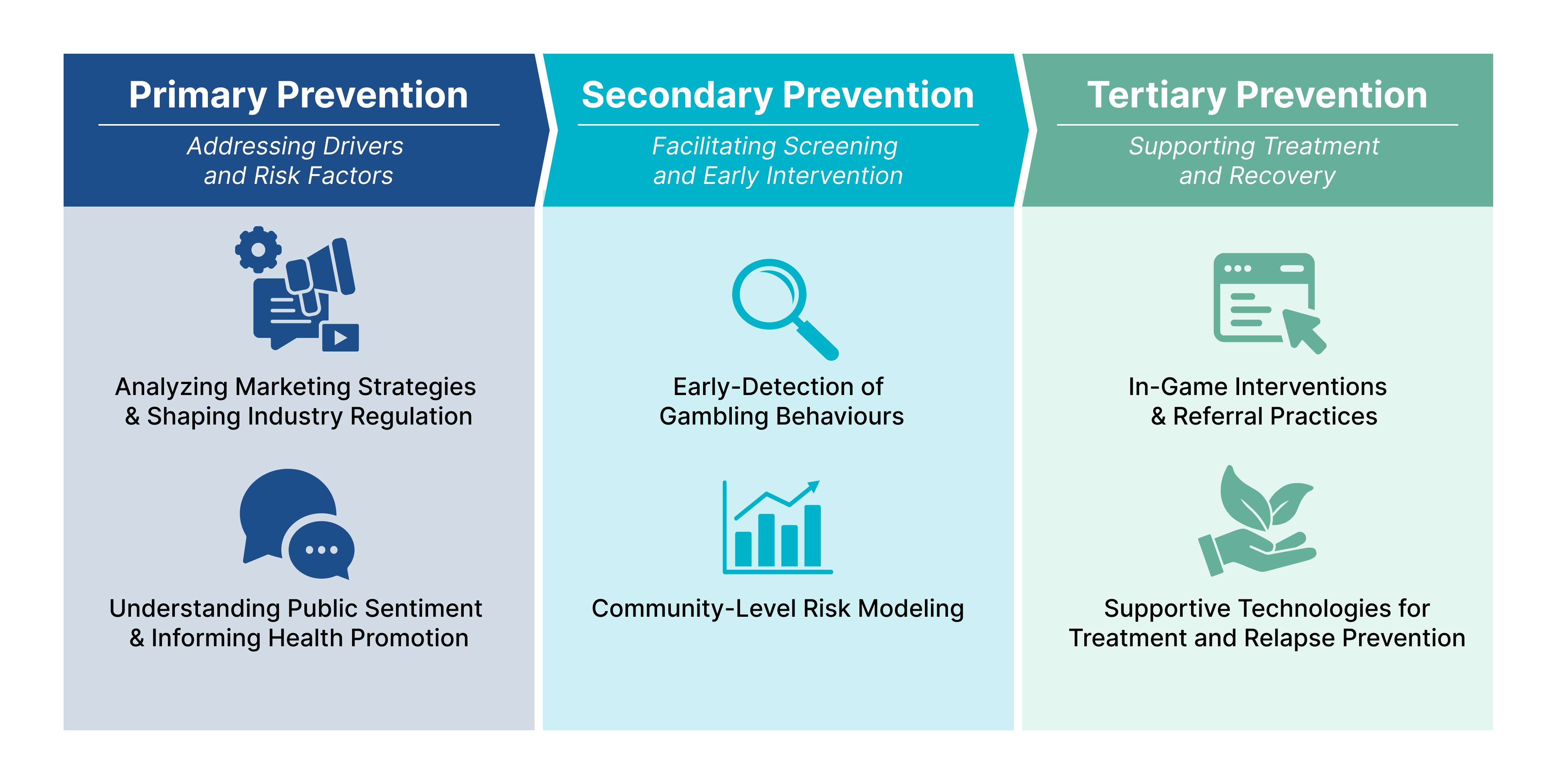}
    \vspace{-5mm}
    \caption{\textbf{Technological interventions in online gambling prevention.} This figure highlights a population health approach with three tiers: primary prevention focuses on risk factor analysis and public health promotion; secondary prevention includes early detection of gambling behaviors and risk modeling; and tertiary prevention addresses treatment and recovery with in-game interventions and relapse prevention technologies.}
    \label{fig:intervention points}
\end{figure*}

\subsection{Prevention Strategies Enabled by AI}
AI-based methods show potential for reducing the public health burden of problem gambling via a structured prevention framework, mirroring general public health models ({Figure \ref{fig:intervention points}}). 

{\bf Primary Prevention.} Primary prevention addresses root causes and broad risk factors before problems arise \citep{kisling2023prevention}. In online gambling, this includes regulating commercial determinants and raising public awareness \citep{petry2018policy}. Multimodal foundation models can analyze text, images, and videos to evaluate marketing tactics—such as how prominently responsible gambling messages appear on advertisements or if certain imagery targets underage audiences. These AI-driven audits foster accountability in advertising, promote clear disclosures, and facilitate more informed public dialogue. Automated sentiment analysis on social media can further capture public perceptions, fueling data-driven efforts to refine prevention campaigns \citep{olawade2023using,lim2023artificial}.

{\bf Secondary Prevention.} Secondary prevention aims to identify and support individuals or communities at higher risk \citep{kisling2023prevention}. Existing work leverages AI-based analyses of user engagement, demographic data, and environmental exposures to pinpoint groups susceptible to developing problem gambling \citep{booth2021affected,monreal2023preventive}. Additional data sources, including social media sentiment on online gambling and demographic trends in site visits, betting frequencies, and wager amounts, may help identify emerging patterns and inform tailored community prevention efforts.

{\bf Tertiary Prevention.} For individuals already exhibiting harmful gambling behavior, tertiary prevention focuses on limiting further damage and fostering recovery \citep{kisling2023prevention}. Screening tools combining machine learning and user engagement metrics (e.g., deposit variability, multiple visits) can predict self-exclusion or problematic gambling with moderate-to-high accuracy \citep{finkenwirth2021using,auer2023using,murch2023using}. For example, {one Canadian study} \citep{finkenwirth2021using} employed 20 betting variables related to gambling frequency, intensity, and variability to develop predictive models for self-exclusion, reporting area under the curve (AUC) values ranging from 0.65 to 0.76. Another {Canadian investigation} \citep{murch2023using} combined user engagement and demographic data, achieving AUC values as high as 0.84 when predicting high-risk behaviors. Such findings have prompted platforms to consider offering personalized feedback, deposit-limit reminders, or prompts to seek help \citep{auer2023impact}. Meanwhile, AI-enabled chatbots or telehealth tools can deliver ongoing monitoring and individualized guidance, further reducing barriers to care.

\begin{figure}[t]
    \centering
    \includegraphics[width=\linewidth]{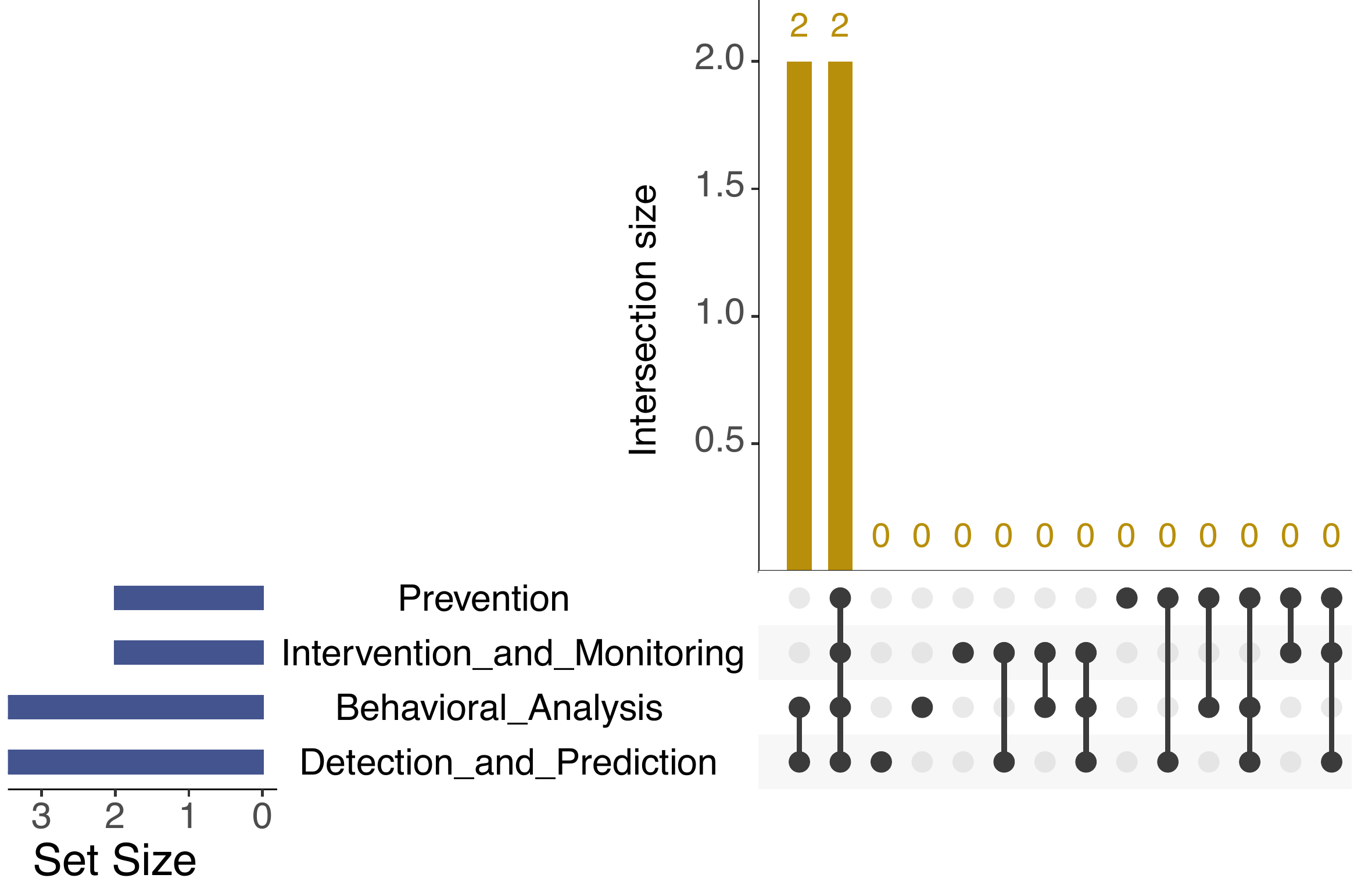}
    \caption{\textbf{UpSet plot illustrating the intersections of requirement vectors across AI application categories.} The vectors are defined as follows: \textit{Detection and Prediction} (\(V_{\mathrm{DP}} = \{1, 0, 1, 0, 0, 1, 0, 1\}\)), \textit{Behavioral Analysis} (\(V_{\mathrm{BA}} = \{1, 0, 1, 0, 0, 1, 0, 1\}\)), \textit{Intervention and Monitoring} (\(V_{\mathrm{IM}} = \{1, 0, 1, 0, 0, 0, 0, 0\}\)), and \textit{Prevention} (\(V_{\mathrm{P}} = \{1, 0, 1, 0, 0, 0, 0, 0\}\)). A size of two for intersections indicates that two requirements co-occur across two categories. A size of zero means that no category combination includes those specific requirements simultaneously. The plot highlights the consistent inclusion of demographic ($r_1$) and behavioral determinants ($r_3$), while systemic factors like policy engagement ($r_7$) and fairness ($r_8$) are underrepresented, particularly in \textit{Intervention and Monitoring} and \textit{Prevention}.}
    \label{fig:upset}
\end{figure}

\subsection{Limitations in AI for Problem Gambling}
Many studies still rely on self-reported data, such as the Problem Gambling Severity Index (PGSI), which is susceptible to recall and social desirability biases \citep{murch2024establishing}. In addition, results based on a single online gambling platform may not capture a user’s broader gambling behaviors, including land-based activities \citep{hopfgartner2024using}. Sample sizes vary greatly, as does the granularity of available data, leading to heterogeneous reporting standards and difficulties in comparing results across different operators or jurisdictions. Evaluation metrics also lack uniform definitions, with area under the curve (AUC), accuracy, precision, and recall frequently used but differently interpreted \citep{murch2024comparing}. Furthermore, co-morbid conditions (e.g., mental health issues) and personal financial factors often remain unaccounted for, creating gaps in the contextual understanding of gambling harm. As AI tools become more prevalent, the sector must address these limitations through larger, more representative datasets, cross-platform validation, and consensus on robust evaluation protocols. Only then can AI fulfill its potential to guide ethical, data-driven strategies for mitigating gambling harm.

Moreover, to provide a broad synthesis of current knowledge and challenges in this evolving field, we propose a structured evaluation of the extent to which key requirements are addressed by various AI application categories in online gambling. Let $R = \{r_{1}, r_{2}, \ldots, r_{8}\}$
denote the set of eight requirements identified from the literature, comprising:

\begin{enumerate}
\footnotesize
    \item $r_{1}$: Demographic Risk Factors
    \item $r_{2}$: Social Determinants
    \item $r_{3}$: Behavioral Determinants
    \item $r_{4}$: Health Determinants
    \item $r_{5}$: Commercial Determinants
    \item $r_{6}$: Real-world Player Data Access
    \item $r_{7}$: Policy Engagement
    \item $r_{8}$: Algorithmic Fairness and Bias
\end{enumerate}

Each AI application category—\textit{Detection and Prediction}, \textit{Behavioral Analysis}, \textit{Intervention and Monitoring}, and \textit{Prevention}—is represented by a binary vector $V_{A} = \{v_1, v_2, \ldots, v_8\}$,
where \(v_i = 1\) indicates that the requirement \(r_i\) is addressed by the existing literature that focused on this specific application, and \(v_i = 0\) indicates it is not. For instance, the vector for \textit{Detection and Prediction} is 
\[
V_{\mathrm{DP}} = \{\,1,\;0,\;1,\;0,\;0,\;1,\;0,\;1\},
\]
reflecting its emphasis on demographic, behavioral, data access, and fairness-oriented components, while placing less priority on social and health determinants. 
Using these representations, an UpSet plot (Figure~\ref{fig:upset}) visualizes the intersections of requirements across these categories. The analysis reveals consistent attention to demographic and behavioral determinants across all application categories, highlighting their foundational role in addressing gambling-related harm. However, systemic considerations like Policy Engagement ($r_7$) and Algorithmic Fairness and Bias ($r_8$) are underrepresented, particularly in \textit{Intervention and Monitoring} and \textit{Prevention}. 

The requirement mapping highlights critical patterns across AI applications in online gambling. Demographic ($r_1$) and behavioral ($r_3$) determinants are consistently prioritized, reflecting their central role in identifying and addressing problem gambling. However, systemic factors such as policy engagement ($r_7$) and algorithmic fairness ($r_8$) are often overlooked, despite their importance for ensuring ethical and regulatory compliance. Moreover, while data access ($r_6$) is essential for \textit{Detection and Prediction} and \textit{Behavioral Analysis}, its limited presence in \textit{Intervention and Monitoring} and \textit{Prevention} presents challenges in acquiring granular, real-time data. Connecting application categories with their requirement vectors exposes these gaps and enables targeted, ethically grounded advancements. The following section elaborates on how \textit{Generative AI and Foundation Models} can tackle these challenges and redefine harm-reduction strategies.

\section{The Potential of Generative AI in Problem Gambling}

This section presents six use cases illustrating how Generative AI and Foundation Models provide integrative and innovative solutions that bridge gaps across disciplines. While previous AI efforts have made strides in addressing these challenges, the advanced capabilities of these models enable more adaptive, scalable, and precise approaches, offering transformative potential for behavioral health and gambling studies.

\subsection{Synthetic Player Profiles}
\label{sec:syn}

Real-world datasets for online gambling research are notoriously difficult to access due to privacy, legal, and proprietary constraints \citep{perrot2022development,jach2024identification,drosatos2018empowering,hopfgartner2024using,gainsbury2015online}. As a result, researchers and practitioners frequently lack the scope to develop robust harm-reduction strategies. By employing large-scale foundation models and modern generative AI, it is now possible to craft \emph{Synthetic Player Profiles} that authentically mirror real gambling behaviors without revealing any individual’s identity \citep{harris2018case}.

Instead of relying on extensive real-world datasets, which are often scarce or inaccessible, we propose using a small anonymized or aggregated collection $\{\mathbf{V}_1,\dots,\mathbf{V}_m\}$ to \emph{calibrate} a foundation model $M$ with parameters $\Theta$. Through a suitable loss function (e.g., negative log-likelihood), $M$ approximates $p(\mathbf{V})$ by learning $q_{\Theta}(\mathbf{V})$ over key behavioral features---such as deposit amounts, session durations, chat text, and self-exclusion flags \citep{shanahan2023role,wang2023rolellm,park2024ai}. Once trained, $M$ produces synthetic profiles $\mathbf{V}^*\sim q_{\Theta^*}(\mathbf{V})$, capturing realistic correlations while preserving anonymity. Privacy mechanisms (e.g., $k$-anonymity, differential privacy) and a rigorous validation pipeline ensure that these synthetic records remain secure and reliable. Thus, even with limited or partial real data, researchers and operators can generate lifelike player trajectories for testing new interventions, policy changes, or predictive models without exposing actual user information.

Transformer-based models, Generative Adversarial Networks (GAN) variants \citep{goodfellow2014generative}, and diffusion-based techniques \citep{yang2023diffusion} can further integrate text logs, biometric signals, or mobile data, creating multifaceted synthetic records. Similar methods have been used in healthcare (e.g., blending clinical notes and diagnostic images) \citep{moor2023foundation,giuffre2023harnessing} and finance (e.g., simulating trader actions) \citep{zuo2024reinforcement}, enabling research free of confidentiality breaches. By aligning generation strategies with the unique traits of regional betting cultures, researchers can reveal subtle risk patterns that standardized models might miss, opening opportunities for more focused and relevant interventions.

Still, careful audits are vital to avoid biases or loss of minority patterns \citep{perrot2022development,gainsbury2015online}. Fairness checks and federated learning can guard original datasets and validate that the synthetic data genuinely reflects real-world distributions \citep{drosatos2018empowering}. Meanwhile, cognitive architectures like CoALA \citep{sumers2023cognitive} and multi-agent simulations \citep{park2023generative} allow these synthetic actors to \textquotesingle{}bet\textquotesingle{}, respond to outcomes, and deploy responsible gambling tools, all with memory-driven and adaptive behavior. Researchers can even simulate real-time chat threads or forum interactions \citep{ghaharian2023applications,moor2023foundation} to capture conversational nuances, including emojis or full natural language. By providing a safe yet realistic testing ground, Synthetic Player Profiles pave the way for large-scale experimentation in online gambling studies. 

\begin{figure}[ht]
    \centering
    \includegraphics[width=\linewidth]{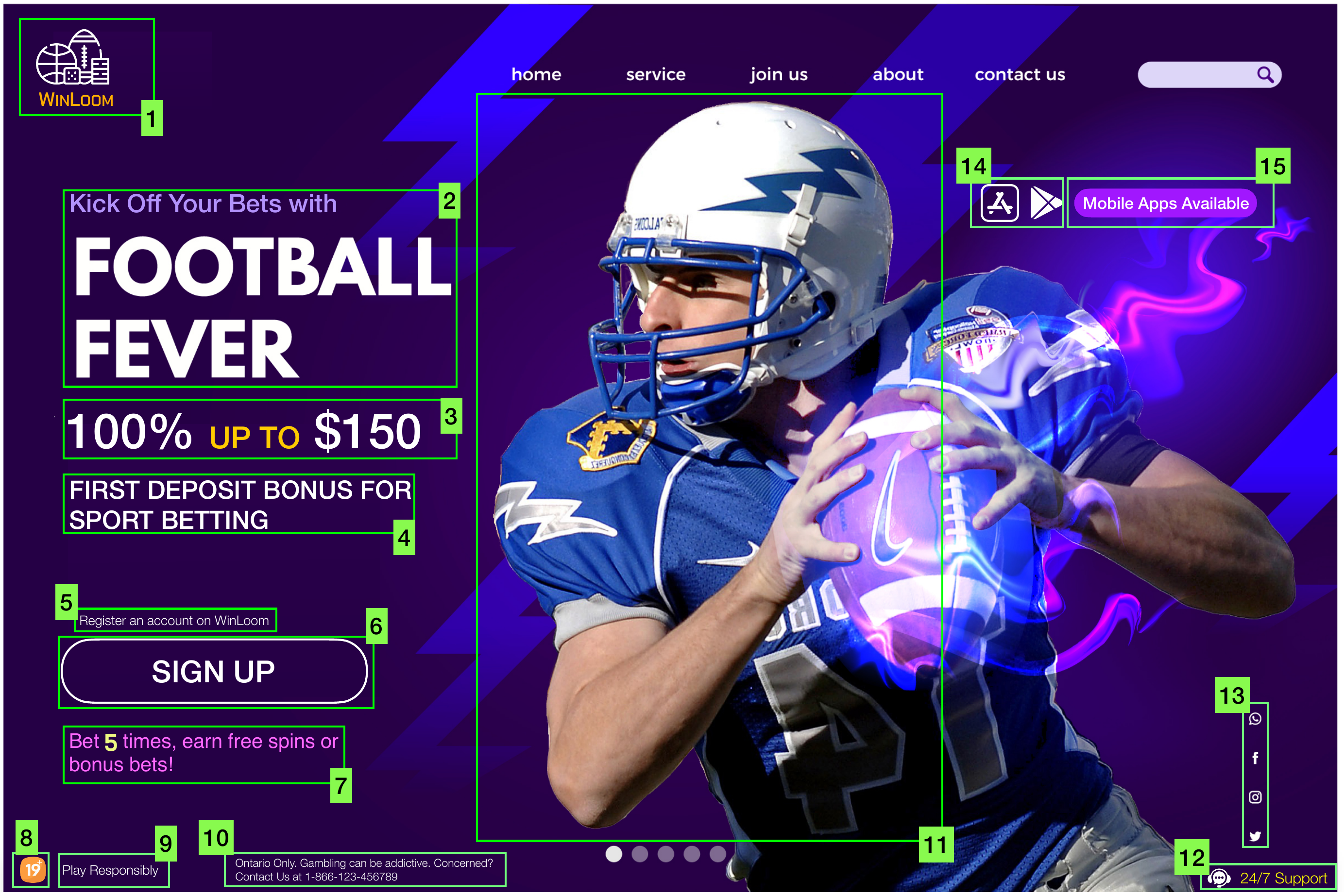}
    \vspace{-2em} 
    \begin{tcolorbox}[colback=blue!1, colframe=black!30, arc=.2mm, boxrule=0.1mm, left=.5mm, right=1mm, top=1mm, bottom=.5mm, width=\linewidth]
    \ttfamily\scriptsize 
    1. {\textcolor{black}{Brand Recognition/Logo:}} {\textcolor{black!70}{Normalizes the brand for underage viewers.}} \\
    2. {\textcolor{black}{Sports Tie-In:}} {\textcolor{black!70}{Links betting to popular sports, attracting teens.}} \\
    3. {\textcolor{black}{Promotional Enticement:}} {\textcolor{black!70}{Portrays \textquotesingle{}free money\textquotesingle{}, skewing risk perception.}} \\
    4. {\textcolor{black}{Signup Bonus:}} {\textcolor{black!70}{Lowers cost barriers, tempting younger users.}} \\
    5. {\textcolor{black}{Direct Call to Action (\textquotesingle{}Register an account\textquotesingle{}):}} {\textcolor{black!70}{Prompts impulsive sign-ups.}}\\ 
    6. {\textcolor{black}{Immediate CTA (\textquotesingle{}Sign Up\textquotesingle{} Button):}} {\textcolor{black!70}{Prompts hasty clicks instead of caution.}} \\
    7. {\textcolor{black}{Loyalty Mechanic (\textquotesingle{}Bet 5 times\textquotesingle{}):}} {\textcolor{black!70}{Promotes repeated betting and addiction.}} \\
    8. {\textcolor{black}{Legitimization (19+ Icon):}} {\textcolor{black!70}{Highlights the legal age, but easily ignored.}}\\
    9. {\textcolor{black}{Compliance Obfuscation:}} {\textcolor{black!70}{Often unnoticed, minimal protection.}}\\ 
    10. {\textcolor{black}{Brief Warning (\textquotesingle{}Gambling can be addictive\textquotesingle{}):}} {\textcolor{black!70}{Overshadowed by bold ads.}} \\
    11. {\textcolor{black}{Aspirational Athlete Imagery:}} {\textcolor{black!70}{Glamorizes gambling via sports heroism.}} \\
    12. {\textcolor{black}{24/7 Support Badge:}} {\textcolor{black!70}{Encourages around-the-clock engagement.}} \\
    13. {\textcolor{black}{Social Media Icons:}} {\textcolor{black!70}{Stay connected with tech-savvy youth.}} \\
    14. {\textcolor{black}{App Store Icons:}} {\textcolor{black!70}{Remove access barriers, even for underage users.}} \\
    15. {\textcolor{black}{Mobile Apps Prompt:}} {\textcolor{black!70}{Promotes on-the-go betting, increasing play risk.}}
\end{tcolorbox}

   \caption{
\textbf{Generative AI-Driven Analysis of Marketing Strategies in Online Gambling Advertisement}.  
An online sports-betting advertisement analyzed using generative AI (\texttt{GPT-4O}). Marketing strategies were detected from visual and textual elements, annotated numerically, and listed below. Model parameters: \texttt{temperature = 0.2}, \texttt{max tokens = 950}, \texttt{top-p = 0.9}, \texttt{frequency penalty = 0.2}, \texttt{presence penalty = 0.0}.
}
    \label{fig:online-gambling-ad}
    \vspace{-5mm}
\end{figure}

\subsection{Responsible Marketing}

The gambling industry increasingly deploys sophisticated, AI-driven marketing strategies to attract and retain customers, raising significant concerns about potential harms as a commercial determinant of health \citep{kshetri2024generative,singer-2024,Dunlop-2021}. These strategies leverage targeted advertising, personalized content, and persuasive messaging via digital channels (e.g., social media, email, in-app notifications), often normalizing gambling as both accessible and glamorous \citep{pettorruso2021transition,baek2023digital,rodgers2021themed}. Such aggressive tactics, including gamified promotions and immersive content, can disproportionately expose underage and vulnerable individuals, exacerbating public health challenges.

Multimodal AI and foundation models, including generative AI and vision-language models (VLMs) \citep{zhu2023minigpt}, offer a powerful means of transforming these practices. They can synthesize and analyze text, images, and videos to detect child-directed language or visuals in ads, allowing regulators to enforce stricter compliance standards \citep{campbell2022preparing,bjorseth2021effects}. One promising application involves scrutinizing influencer content on social platforms \citep{kim2020multimodal}, flagging gambling promotions lacking proper disclosure \citep{pitt2024young,warraich2024fda}. Similarly, tools used for misinformation detection in political campaigns \citep{chen2024combating,kuznetsova2025generative,kumar2024feature} could be adapted to evaluate the factual accuracy and ethical framing of gambling advertisements. In line with broader advertising insights \citep{grewal2024generative,zhang2021multimodal}, AI-based audits can also uncover manipulative or stereotype-reinforcing elements \citep{shumanov2022using,liu2022artificial}, ensuring that marketing does not glorify gambling as a fast path to financial security.

\medskip

To illustrate these concepts more concretely, we define an AI-driven function $f_{\Theta}:(\mathcal{M},\mathcal{U})\rightarrow \mathcal{R}$ that assigns a risk or relevance score $r \in \mathcal{R}$ to marketing materials aimed at specific user groups. Let $\mathcal{M}=\{m_1,\dots,m_k\}$ be marketing items (e.g., text, images, videos) and let $\mathcal{U}=\{u_1,\dots,u_n\}$ denote user segments defined by demographics, behaviors, and social ties. In practice, $f_{\Theta}$ combines VLMs to process multimodal content \citep{zhu2023minigpt} and graph neural networks (GNNs) to account for relational data \citep{wu2020comprehensive,bhadra2023graph}. Once trained on labeled examples (e.g., ads deemed high-risk), $f_{\Theta}$ flags harmful content and recommends interventions (e.g., restricting ad distribution, altering imagery, or providing disclaimers). This pipeline can incorporate privacy safeguards (e.g., data anonymization) and continuous validation (e.g., adherence to ethical marketing guidelines) to ensure transparency and social responsibility.

As shown in Figure~\ref{fig:online-gambling-ad}, generative AI can annotate suspicious elements ranging from exaggerated reward claims to youth-targeted graphics, thereby illustrating how $f_{\Theta}$ might assign higher risk scores to unethical or misleading promotions. Another emerging technique features GNNs that model intricate relationships among users, behaviors, and content \citep{bhadra2023graph}. By mapping how a promotion spreads through a user’s social network, these models can forecast cascading effects and dynamically adjust campaigns to reduce harm. When enhanced with emotion recognition layers, they can further detect ads that exploit anxiety or overconfidence, proactively flagging them for immediate review.

\subsection{Personalized Behavioral Interventions}

Generative AI heralds a paradigm shift in delivering personalized behavioral interventions for online gambling, moving from static approaches to dynamic, adaptive strategies \citep{baig2024generative,yang2024chatdiet,geng2022recommendation}. Large Language Models (LLMs) and other generative methods enable real-time customization of prompts, nudges, and supportive messages based on user data such as spending habits, session durations, and even emotional tone \citep{aggarwal2023artificial,an2024scoping,nie2024llm}. This mirrors personalized marketing, yet applied to health and well-being \citep{lee2024developing}, allowing interventions to be generated \emph{on the fly} at critical moments and potentially halting escalation into problematic gambling behaviors. Let $\mathcal{D}=\{d_1,\dots,d_m\}$ represent key data streams (e.g., betting amounts, session timestamps, chat logs), and let $\mathcal{I}=\{i_1,\dots,i_k\}$ be a set of intervention strategies (e.g., nudges, cooldown reminders, self-reflection prompts). We define an AI-driven mapping $g_{\Theta}:\mathcal{D}\rightarrow\mathcal{I}$, where the generative model’s parameters $\Theta$ are calibrated using real or \emph{synthetic data} to ensure diverse scenarios while preserving privacy. Once trained on historical gambling patterns and labeled intervention outcomes, $g_{\Theta}$ delivers personalized, context-aware content in real time, with safeguards like data anonymization and ethical review boards mitigating potential misuse.

The personalization deepens when sentiment analysis and contextual data are integrated, allowing the system to react differently to frustration, overconfidence, or distress \citep{nepal2024mindscape,an2024scoping,maher2020physical,subramanian2024graph}. For instance, a user on a losing streak might receive empathetic messages encouraging a break, while a winning user might see reminders about responsible wagering limits. This Generative AI-based adaptability parallels advancements in mental health apps, where contextual prompts foster self-reflection \citep{kirk2024benefits,moor2023foundation}. One promising extension is to use digital \textquotesingle{}micro-rewards\textquotesingle{}, which gently incentivize players to follow recommended limits or take breaks, reinforcing positive behaviors without relying on intrusive tactics. Over time, this targeted personalization fosters a supportive environment in which users feel guided and motivated to make healthier gambling choices.

\subsection{Gamified Recovery Tools}

Generative AI offers powerful new possibilities for developing interactive, gamified recovery tools tailored for individuals addressing problem gambling \citep{stade2024large,obradovich2024opportunities}. These tools move beyond traditional approaches by providing engaging, personalized experiences that tackle the unique challenges of gambling addiction \citep{cheng2023gamification}. For example, AI-generated simulations can mimic real-life gambling scenarios within a safe, controlled digital environment, enabling users to practice impulse control and decision-making skills under realistic triggers \citep{sezgin2024behavioral}. This strategy parallels the use of serious games in education and health, where immersive learning environments reinforce critical competencies.

In addition to simulations, generative AI can develop gamified exercises that reward positive behaviors linked to recovery, such as adhering to self-exclusion programs or attending therapy sessions \citep{cheng2023gamification}. These elements boost motivation and adherence, both vital for maintaining progress. Moreover, AI can produce tailored narratives and videos that highlight others’ recovery stories, instilling a sense of community, hope, and relatable success \citep{sezgin2024behavioral,arenas2022therapeutic,nye2023efficacy}. Let $\mathcal{S}=\{s_1,\dots,s_n\}$ represent a user's state (e.g., emotional cues, triggers, progress metrics), and let $\mathcal{M}=\{m_1,\dots,m_k\}$ denote gamified modules (e.g., simulations, reward systems, narrative exercises). We define a generative mapping $h_{\Theta}:\mathcal{S}\to \mathcal{M}$, where the parameters $\Theta$ are learned from real or \emph{synthetic data} (see Section~\ref{sec:syn}) to preserve privacy. Once trained, $h_{\Theta}$ personalizes each module in real time, adjusting difficulty or content based on user feedback \citep{sezgin2024behavioral,nye2023efficacy}. Incorporating wearable-sensor data (e.g., heart rate) into $\mathcal{S}$ could enhance early detection of emotional distress, prompting instant delivery of coping strategies.

Generative AI has already been leveraged in mental health to create personalized images for emotional management \citep{sezgin2024behavioral}, and similar logic applies to gambling recovery: dynamic content and adaptive reward systems can drive sustained engagement \citep{li2023systematic,obradovich2024opportunities}. AI chatbots can further complement these modules, offering guidance and real-time interventions between human therapy sessions. The integration of simulations, gamified exercises, and on-demand support delivers a comprehensive and data-driven approach to achieving lasting recovery. 

\begin{figure*}
    \centering
    \includegraphics[width=0.85\linewidth]{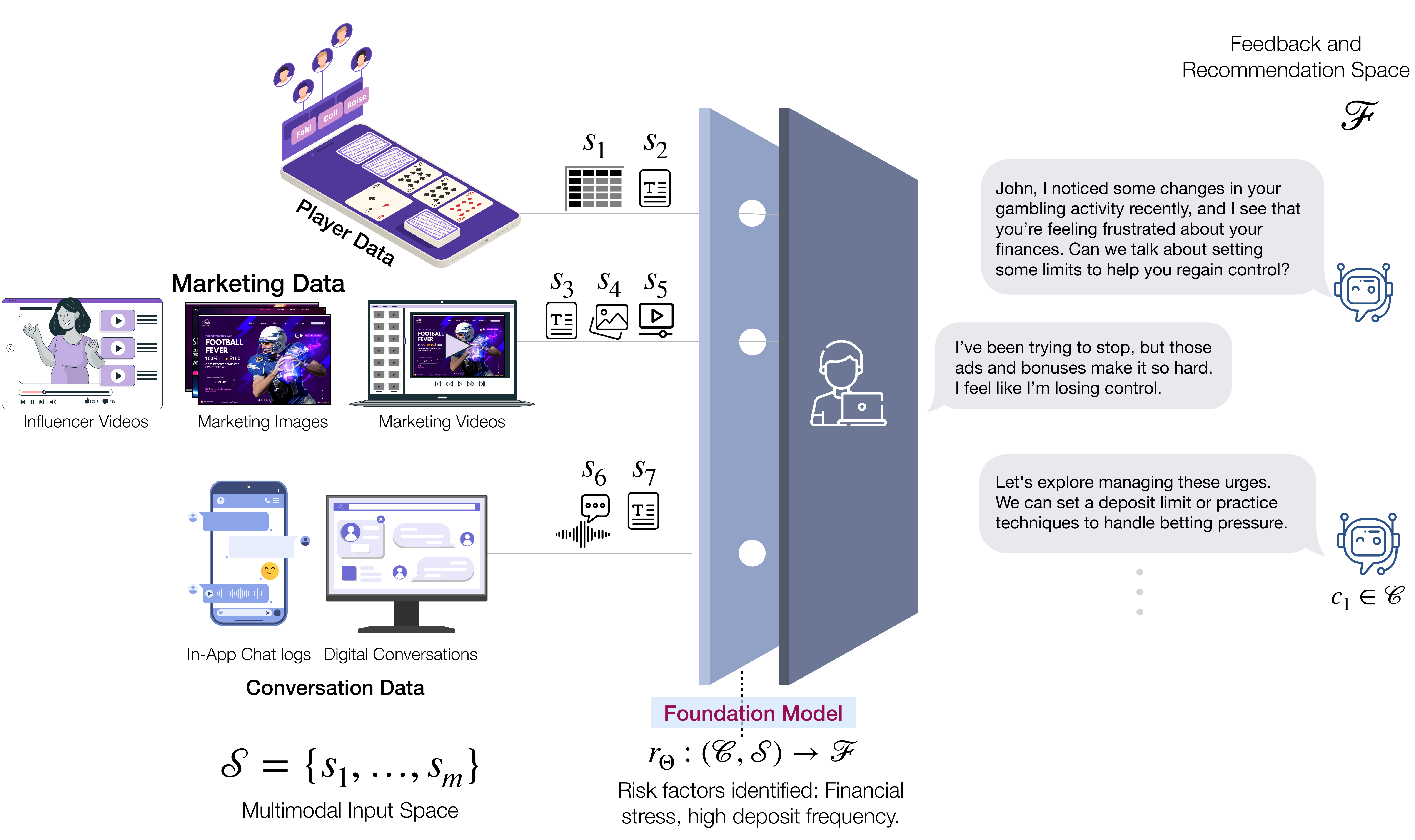}
    \caption[Multimodal Foundation Model for Problem Gambling Counseling]{\textbf{Multimodal Foundation Model for Problem Gambling Counseling.} This figure illustrates the integration of multimodal data (\(\mathcal{S} = \{s_1, \dots, s_m\}\)), including player behaviors, marketing influences, and conversational cues, into a foundation model (\(r_{\Theta}\)). The model processes these inputs to generate actionable insights (\(\mathcal{F}\)) that guide counselors (\(\mathcal{C}\)) in providing tailored interventions. The counselor (\(c_1 \in \mathcal{C}\)) interacts with the model’s recommendations to address specific client concerns, such as identifying risk factors (e.g., financial stress, high deposit frequency) and proposing strategies like deposit limits or behavioral exercises.}

    \label{fig:counceling}
\end{figure*}

\subsection{AI-Driven Counselor Training and Decision Support}

The rising complexities of online gambling demand innovative approaches to counselor training and decision support, and multimodal AI coupled with foundation models offers a transformative solution \citep{shen2024large,li2023systematic}. One particularly promising use case is the creation of AI-driven training platforms that simulate realistic counseling scenarios for diverse behavioral health issues \citep{shen2024large,steenstra2024virtual}. For instance, a virtual client could present with the intertwined challenges of financial distress, strained relationships, and comorbid mental health conditions, incorporating voice tone, facial expressions, and text-based interactions to capture the nuance of real sessions \citep{stade2024large,magill2022technology}. This mirrors systems in addiction counseling (e.g., substance abuse \citep{steenstra2024virtual}) and mental health (e.g., depression or anxiety \citep{stade2024large}), where trainees practice techniques like Cognitive Behavioral Therapy (CBT) or motivational interviewing and receive instant, targeted feedback.

Beyond training, AI provides vital clinical decision support across multiple domains. By analyzing gambling transactions, online behaviors, and communication data, clinicians gain a holistic view of client patterns \citep{shen2024large,maslej2023out}. Similar analytics have been applied to eating disorders (e.g., tracking food logs, mood trends) \citep{li2023systematic} and chronic pain management (e.g., monitoring activity levels, medication use) \citep{abedi2024artificial}, all with the goal of enabling informed, personalized interventions \citep{stade2024large}. AI-powered conversational agents can also serve as virtual peer counselors, helping practitioners process complex cases or manage emotional stress \citep{stade2024large,shen2024large,xi2023rise}. This mirrors trauma or family therapy settings, where virtual peers offer real-time guidance on handling sensitive topics \citep{xi2023rise,abedi2024artificial}. Let $\mathcal{C}=\{c_1,\dots,c_p\}$ be counselor trainees or practicing clinicians, and let $\mathcal{S}=\{s_1,\dots,s_m\}$ represent multimodal simulation elements (e.g., text dialogues, vocal tones, facial cues, or virtual environments). We define a generative mapping $r_{\Theta}:(\mathcal{C},\mathcal{S}) \rightarrow \mathcal{F}$, where $\mathcal{F}$ is a feedback or recommendation space (e.g., performance metrics, evidence-based treatment steps). Parameters $\Theta$ are learned from real or \emph{synthetic data} that reflect various behavioral health scenarios. Once trained, $r_{\Theta}$ dynamically adapts each simulation to individual counselor needs, suggesting tailored techniques (e.g., CBT vs.\ motivational interviewing) or even highlighting relevant ethical considerations (Figure \ref{fig:counceling}).   

Hyper-realistic virtual environments further expand these capabilities by recreating specific situations such as online casinos \citep{magill2022technology}, helping counselors grasp the triggers and temptations their clients face. Comparable setups apply to social anxiety, where users virtually encounter stress-inducing social scenarios. While these technologies offer substantial promise, it remains essential to balance innovation with ethics \citep{althoff2016large,xi2023rise}, ensuring user safety, fairness, and respect for privacy. Augmenting rather than replacing human expertise, AI-driven counselor training and decision support have the potential to transform the quality and accessibility of behavioral health services.

\subsection{Scenario Modeling for Policy Development}

Generative AI enables a proactive and data-driven approach to policy-making by creating detailed scenario models and simulations \citep{cao2024llm,barnett2024simulating}. Rather than merely reacting to existing issues, policymakers can explore \textquotesingle{}what-if\textquotesingle{} scenarios in gambling regulation, such as changing advertising standards, adjusting self-exclusion rules, or altering tax structures, and instantly gauge their effects on addiction rates, economic outcomes, and social dynamics \citep{barnett2024simulating,tyler2023ai}. Similar modeling applies to broader public policy contexts like environmental measures or public health, where AI can integrate large datasets (e.g., demographics, economic indicators) to provide intuitive visual reports and interactive forecasts. In the context of this use case, let $\mathcal{P}=\{p_1,\dots,p_z\}$ denote a set of policy options (e.g., advertising limits, tax incentives), and let $\mathcal{V}=\{v_1,\dots,v_m\}$ be relevant variables (e.g., player behaviors, demographics). We define $S_{\Theta}:(\mathcal{P}, \mathcal{V})\to \mathcal{O}$, where $\mathcal{O}$ captures model outcomes (e.g., projected addiction rates, revenue forecasts). Once $S_{\Theta}$ is trained—potentially leveraging \emph{synthetic data} for privacy and coverage (see Section~\ref{sec:syn})—policymakers can manipulate $\mathcal{P}$ in real time, generating scenario-driven insights to inform robust, forward-thinking regulations. One promising extension is bridging cross-border regulatory differences via a single integrated model capable of unifying data from diverse jurisdictions.

This ability to produce complex, dynamic simulations democratizes access to actionable insights, enabling non-technical stakeholders to engage meaningfully with policy decisions \citep{cao2024llm}. Experimenting with different policy levers and examining likely outcomes helps governments, industry leaders, and advocacy groups anticipate unintended consequences and refine strategies. AI-driven scenario modeling turns policy-making into a dynamic and forward-thinking process, proving safer and more accountable gambling ecosystems.

\section{Ethical and Technical Challenges}

While AI technology presents promising opportunities to reduce online gambling harm, practitioners must contend with a range of ethical and technical challenges in deploying these systems. First, \textbf{validation} becomes significantly more complex when dealing with large, versatile AI models. Unlike conventional software designed for narrow tasks, these models can be repurposed in unforeseen ways, complicating efforts to certify performance across multiple domains \citep{mitchell2019model}. For instance, an algorithm trained primarily for risk detection could be co-opted to nudge players toward higher spending, necessitating rigorous oversight and continual monitoring.

A related concern is \textbf{verification}. Multimodal models that combine textual, behavioral, and biometric data can be daunting for public health practitioners or non-technical stakeholders to evaluate \citep{holstein2019improving}. For example, an AI-driven system might flag excessive betting behaviors while also producing specialized visual or statistical justifications. Verifying the correctness of these outputs, potentially involving large datasets, fluid user contexts, or proprietary embedding layers, often requires cross-disciplinary reviews and tailored explainability tools \citep{bender2021dangers}.

\textbf{Social biases} pose an equally urgent problem. Large-scale models risk perpetuating or amplifying unfair outcomes, especially for minority groups underrepresented in training data \citep{koenecke2020racial}. In the gambling context, this could manifest as AI interventions that overlook certain demographics, failing to provide timely harm-reduction measures. Frequent audits of model outputs and collaboration with domain experts who understand local cultural nuances are essential to detect and mitigate such biases.

\textbf{Privacy} remains a pressing issue as well. Foundation models often rely on massive corpora of user-generated content, which may inadvertently embed personally identifiable data \citep{gupta2023chatgpt}. Even anonymized gambling logs can be susceptible to re-identification attacks, especially if combined with external data sources. Moreover, \textquotesingle{}prompt attacks\textquotesingle{} or \textquotesingle{}jailbreak\textquotesingle{} exploits can lure a model into revealing sensitive user information despite safeguards \citep{weidinger2021ethical}. Robust governance measures, including data minimization, encryption, and real-time anomaly detection, are crucial to protect users in this high-risk environment.

Finally, the \textbf{scale} of modern AI models raises both environmental and economic concerns. Training and updating large-scale systems can be computationally expensive, resulting in substantial carbon footprints and elevated operational costs \citep{patterson2022carbon}. Gambling platforms that rely on frequently updated models must balance sophistication with sustainability to avoid prohibitive resource demands.

In addition to these technical hurdles, practitioners must also address overarching \textbf{ethical considerations} related to data collection, transparency, and industry accountability. For instance, re-identification risks could escalate when operators link user betting data (e.g., for limit-setting) with browsing histories to deliver personalized gambling ads. Deciding which data are gathered and how it is stored becomes central to preventing downstream harms \citep{price2019privacy}. Likewise, clear communication about how data are collected, models are developed, and technologies are deployed is vital for fostering user autonomy and community trust \citep{fisher2022priorities}. Strengthening multi-stakeholder governance boards, including healthcare professionals, ethicists, and gambling regulators, could help ensure that AI interventions complement, rather than replace, the industry’s duty to address addictive platform features. Although tools such as limit setting, feedback messaging, and service referrals are beneficial, these do not necessarily tackle the root causes of problematic behaviors, highlighting the need for continued vigilance around platform design and industry practices.

\section{Conclusion}

This review highlights the evolving challenges posed by online gambling and the opportunities that artificial intelligence, particularly multimodal generative AI and foundation models, offers for prevention and harm reduction. These technologies enable advanced analytics, adaptive interventions, and realistic policy modeling, expanding the possibilities for responsible gambling research and practice. At the same time, they emphasize the need for safeguarding data privacy, mitigating algorithmic bias, and maintaining accountability to protect user well-being.

The examples discussed in this review demonstrate how AI can enhance our understanding of gambling behaviors, support the development of personalized harm-reduction tools, and inform regulatory strategies. Moving forward, validating these applications across diverse contexts and addressing their limitations will be crucial. Collaboration among researchers, clinicians, policymakers, and technologists is key to ensuring these tools are implemented responsibly and effectively. With careful integration, AI has the potential to redefine prevention, early detection, and treatment strategies for online gambling, supporting meaningful progress in addressing gambling-related harms.

\bibliographystyle{cas-model2-names}
\bibliography{GamblingPerspective.bib}
\end{document}